# Precursor outbursts and superoutbursts in the SU UMa-type dwarf nova NN Camelopardalis

Jeremy Shears, Jerry Foote, William Mack Julian, Tom Krajci, Igor Kudzej, Ian Miller, Etienne Morelle, Richard Sabo, Irina Solovyova, Bart Staels and Tonny Vanmunster

## Abstract

We report photometry of three outbursts of NN Cam in 2007, 2008 and 2009. The 2007 event started with a normal outburst, lasting about 4 days, which was a precursor to a superoutburst lasting at least 13 days. Both the precursor and the superoutburst had an amplitude of 4.9 mag above mean quiescence. Superhumps with a maximum peak-to-peak amplitude of 0.22 mag were detected during the superoutburst with a mean superhump period $P_{sh}$ = 0.07385(56) d. $P_{sh}$ decreased continuously with $dP_{sh}/dt$ = -1.72(23) x $10^{-3}$. We used our measurement to confirm that the shorter of two possible values of $P_{orb}$ reported by another researcher is the correct one, $P_{orb}$ = 0.0717 d. The 2008 outburst was rather poorly observed, although we present evidence that this too may have been a superoutburst. The 2009 event was also a superoutburst, with $P_{sh}$ = 0.07414(44) d, but we could find no evidence for a precursor. From the 2007 and 2009 data, we report a superhump period excess of $\varepsilon$ = 0.030(8) to 0.034(6), which is typical for SU UMa dwarf novae of similar orbital period, and estimate the binary mass ratio $q = M_{wd}/M_{sec} \approx$ 0.11 to 0.17.

## Introduction

Dwarf novae are a class of cataclysmic variable star in which a white dwarf primary accretes material from a secondary star via Roche lobe overflow. The secondary is usually a late-type main-sequence star. In the absence of a significant white dwarf magnetic field, material from the secondary passes through an accretion disc before settling on the surface of the white dwarf. As material builds up in the disc, a thermal instability is triggered that drives the disc into a hotter, brighter state causing an outburst in which the star apparently brightens by several magnitudes [1]. Dwarf novae of the SU UMa family occasionally exhibit superoutbursts which last several times longer than normal outbursts and may be up to a magnitude brighter. During a superoutburst the light curve of an SU UMa system is characterised by superhumps. These are modulations in the light curve which are a few percent longer than the orbital period. They are thought to arise from the interaction of the secondary star orbit with a slowly precessing eccentric accretion disc. The eccentricity of the disc arises because a 3:1 resonance occurs between the secondary star orbit and the motion of matter in the outer accretion disc. For a more detailed review of SU UMa dwarf novae and superhumps, the reader is directed to reference 1.

## History of NN Cam

The discovery circumstances and observational history of NN Cam (=NSV 1485 = SVS 958) have been investigated and described by Denis V. Denisenko [2] of the Space Research Institute (IKI), Russian Academy of Sciences. The star was originally discovered on Moscow photographic plates in 1944 by Meshkova as a variable of unknown type with a magnitude range of 12.9 to >14.9 [3]. Denisenko noted that NN Cam has sometimes been mistaken for another star about 8" to the east, USNO-A2.0 1575-01945880 (R=14.7, B=15.2), including in a recent paper by Khruslov [4]. Khruslov reported a magnitude range of R= 13.2 - 15.6 based on ROTSE data and noted one outburst, leading him to conclude that it is a U Gem type dwarf nova [4]. Denisenko pointed out that the angular resolution of ROTSE is such that the measurements not only include NN Cam and USNO-A2.0 1575-01945880, but also a third star, USNO-A2.0 1575-01945732 [4]. NSVS, which is based on ROTSE data, shows two outbursts [5]:

    JD 2451437 (1999 Sep 17)       R = 14.0 at maximum
    JD 2451521 (1999 Dec 8)        R = 13.2 at maximum (also noted by Khruslov [4])

Denisenko [2] monitored NN Cam with the Russian-Turkish 1.5 m Telescope (RTT 150) on 6 nights between 2007 Aug 26 and Sep 7 and found the star at quiescence varying between 17.2 and 17.8. He reported two possible orbital periods of 0.0717 d or 0.0771 d, being 1 cycle/d aliases, and a position of RA 04h 12 min 26.89 s Dec. +69 deg 29 min 06.4 s (J2000). The outbursting nature of NN Cam and the fact that the orbital period is below the period gap in the orbital period distribution of dwarf novae led him to suggest it is a dwarf nova of the SU UMa family [2].

NN Cam was reported to be in outburst on 2007 Sep 10 by Denisenko and he called for follow up photometry [2]. We report time resolved photometry from this outburst as well as subsequent outbursts in 2008 and 2009.

**Time resolved photometry**

V-band and unfiltered (Clear, "C") time resolved photometry was conducted during the 2007, 2008 and 2009 outbursts using the instrumentation shown in Table 1 and a log of observations is given in Tables 2 to 4. Each observer calibrated their own images by dark-subtraction and flat-fielding and then carried out differential photometry using V-band photometry of comparison stars from AAVSO sequence 1036apk [6]. Given that each observer used slightly different instrumentation, including CCD cameras with different spectral responses, small systematic differences are likely to exist between observers. Where overlapping datasets were obtained, we aligned measurements by different observers by experiment. In this case adjustments of up to 0.07 magnitude were made; in cases where no overlap occurred, no correction was made. However, given that the aim of the time resolved photometry was to investigate periodic variations in the light curve, we consider this not to be a significant disadvantage. Heliocentric corrections were applied to all data.

**The 2007 outburst**

The 2007 outburst light curve is shown in the top panel of Figure 1, which includes our time resolved photometry supplemented with individual data points from the AAVSO International Database. The most striking feature is that there appear to be two outbursts in close succession. The first outburst lasted about 4 days and reached V=12.6, some 4.9 magnitudes above mean quiescence, after which it declined rapidly (1.0 mag/d). The speed of decline, combined to its rather short duration, leads us to conclude this was a normal outburst. Due to the lack of data we cannot say whether the star returned to quiescence at the end of the normal outburst, but 6 days after the last observation at C = 15.5, the star was unexpectedly found by Patrick Schmeer [7] to be in outburst for a second time at v=12.9. The first 6 days of the new outburst corresponded to the plateau phase during which there was a slow fade at ~0.22 mag/d, after which there was a rapid decline. The final return to quiescence was not observed, so the duration is not well constrained, but the star was still well above quiescence at mag 16.1 some 13 days after

Schmeer's detection of the second outburst. Taking into account both outbursts, the total duration was at least 22 days.

We plot expanded views of the time resolved photometry from the outburst in Figure 2. The bottom 3 panels, which contain photometry during the second outburst, clearly show the presence of superhumps, showing this to be a superoutburst, the first such confirmed. The peak-to-peak superhump amplitude is shown in the middle panel of Figure 1. We found that the superhump amplitude increased from 0.14 mag to 0.22 mag during the first two nights of observation (JD 2454363 to 2454365), after which it gradually decreased to 0.08 mag towards the end of the plateau phase. There is evidence that the superhumps subsequently re-grew during the rapid decline.

To study the superhump behaviour, we first extracted the times of each sufficiently well-defined superhump maximum by fitting a quadratic function to the top part of each superhump. Times of 28 superhump maxima were found and are listed in Table 5. We assigned preliminary superhump cycle numbers to these maxima. An analysis of the times of maximum allowed us to obtain the following linear superhump maximum ephemeris:

$$HJD_{max} = 2454363.5574(52) + 0.07385(56) \times E \qquad \text{Equation 1}$$

Thus the mean superhump period was $P_{sh}$ = 0.07385(56) d. The observed minus calculated (O–C) residuals for all the superhump maxima relative to the ephemeris are shown in the bottom panel of Figure 1. The plot clearly shows that the superhump period was decreasing during the superoutburst. The data are consistent with a continuous change in period with $dP_{sh}/dt = -1.72(23) \times 10^{-3}$

Time resolved photometry from the first outburst is shown in the first panel of Figure 2. Small modulations (~ 0.08 mag) are visible, although it was not possible to measure their times of maximum. To investigate the modulations further we carried out a period analysis of the data using the Lomb-Scargle algorithm in the Peranso software [8], having first subtracted the linear trend of the data. The resulting power spectrum (Figure 3) has a multiplicity of 1 cycle/day aliases, the strongest three in order of strength being:

1. 12.972(239) cycles/day          0.0771(14) d
2. 11.972(242) cycles/day          0.0825(15) d
3. 13.971(204) cycles/day          0.0716(11) d

Errors are determined by the Schwarzenberg-Czerny method [9]. We note that signals 1 and 3 are consistent with the two values of $P_{orb}$ proposed by Denisenko [2]. On the other hand none of the signals are consistent with our measured value of $P_{sh}$. Thus we suggest that the declining phase of the normal outburst was modulated with $P_{orb}$.

**The 2009 outburst**

The 2009 outburst light curve is shown in the top panel of Figure 4. The profile is different from the 2007 superoutburst in that there was apparently no preceding normal outburst and the fade during the plateau phase was more gradual (0.06 mag/d). Again the approach to quiescence was not well observed, but the star was still above quiescence some 15 days after detection.

Superhumps were observed throughout the outburst (Figure 5), confirming that this was a superoutburst. The peak-to-peak superhump amplitude is shown in the middle panel of

Figure 3. During the first night (JD 2455137) the superhumps were barely detectable, having an amplitude of ~0.02 mag, suggesting the superoutburst had been caught in its early stages. On the following night they grew from an amplitude of 0.08 mag to 0.14 mag. The superhumps reached a maximum amplitude of 0.26 mag on 2455140, midway through the plateau phase, and subsequently declined to about 0.15 mag, where they remained during the rest of the plateau.

We measured the times of 17 superhump maxima (Table 6) and, as before, fitted these times to a linear ephemeris:

$$HJD_{max} = 2455138.6766(51) + 0.07414(44) \times E \qquad \text{Equation 2}$$

The measured value of $P_{sh}$ = 0.07414(44) d is consistent with that of the 2007 superoutburst.

The O–C residuals relative to this ephemeris are plotted in the bottom panel of Figure 4. Although there was a spread in O-C of about +/- 0.05 superhump cycles, there was no evidence for a continuous change in period as had been seen in 2007.

**The 2008 outburst**

We leave the 2008 outburst until last as it is the least well characterised of the three outbursts. The light curve is plotted in Figure 6, to the same scale as used for the 2007 and 2009 outburst for ease of comparison. The outburst was detected on JD 2454758 at mag 13.2; the previous reported observation was 12 days earlier when the star was in quiescence at mag 17.4. Some 25 days after outburst detection the star was still well above quiescence at magnitude 14.9

Time resolved photometry conducted on the night after discovery (JD 2454759) revealed that the star was declining rapidly (1.2 mag/d) and that there were no obvious modulations in the light curve (Lomb-Scargle analysis of the data also failed to reveal any prominent signals – data not shown). Thus this was almost certainly a normal outburst. Unfortunately there was a gap of 22 days until the next observations on JD 2454781. Expanded views of the photometry on 3 nights from JD 2454781 to 2454784 are shown in Figure 7. Whilst no obvious modulations are apparent to the eye, we note that brightness variations of 0.1 to 0.2 mag are present, which is well above the mean noise level of the individual observations (0.013 mag). To investigate this further we carried out a period analysis of the data in Figure 7 using the Lomb-Scargle algorithm in the Peranso as described above. The resulting power spectrum is shown in Figure 8. The six strongest signals in the spectrum can be divided into 2 groups which are shown in Table 7, where the signals in each group are identified by coloured bars. The three signals in each group are 1 cycle/day aliases of each other. We suggest Group 1 (red bars) represents aliases of $P_{orb}$ since that the strongest 2 signals in this groups are consistent with the two $P_{orb}$ aliases reported by Denisenko [2]. By contrast, the strongest signal in Group 2 (blue bars) is consistent with our values of $P_{sh}$ from 2007 and 2009. If these signals are due to superhumps it would confirm that the star was in superoutburst. The fact that the star was at a similar magnitude over 3 consecutive nights is also consistent with this being part of the plateau phase of a superoutburst.

Unfortunately the 2008 outburst light curve was too poorly sampled to conclude whether there was a normal outburst followed by a superoutburst.

**Discussion**

Our measurements of $P_{sh}$ enable us to rule out the longer of the two possible $P_{orb}$ values proposed by Denisenko [2] (i.e. 0.0771 d) since such value is greater than $P_{sh}$, which would be a most unusual result for an SU UMa system where $P_{sh}$ is normally a few percent longer than $P_{orb}$. Thus we infer that Denisenko's $P_{orb}$ = 0.0717 d is the correct one. Taking this value of $P_{orb}$ and our $P_{sh}$ measurements from the 2007 and 2009 outbursts, 0.07385(56) d and 0.07414(44) respectively, allows the fractional superhump period excess $\varepsilon = (P_{sh} - P_{orb}) / P_{orb}$ to be calculated as 0.030(8) or 0.034(6). Both values are consistent with the range of $\varepsilon$ observed in other SU UMa dwarf novae with similar $P_{orb}$ [10]. Measuring $\varepsilon$ provides a way to estimate the mass ratio $q = M_{wd}/M_{sec}$ of a dwarf nova. Patterson *et al.* [10] developed the empirical relationship $\varepsilon = 0.18\ q + 0.29\ q^2$, which allows us to estimate $q \approx 0.11$ to 0.17.

One of the most intriguing aspects of the 2007 outburst is the observation of a normal outburst preceding the superoutburst. From published superoutburst light curves of SU UMa dwarf novae, it appears that precursor outbursts occur in some, but by no means all such systems. In the thermal-tidal instability model for superoutbursts, a normal outburst acts as a trigger for the superoutburst [11]. Precursor outbursts have been observed in HS 0417+7445 (= 1RXS J042332+745300) [12], GO Com [13], ASAS J224349+0809.5 [14], TV Cor [15], QZ Vir (=T Leo) [16] and VW Hyi [17, 18, 19]. VW Hyi is probably the SU UMa dwarf nova with the best-documented long-term light curve and precursor outbursts are seen during some, but not all superoutbursts [17, 18, 19]. Moreover the time lag between the precursor outburst and the superoutburst is variable: in some cases the precursor and the superoutburst are clearly separated, in others there is a smooth transition from one to the other and in yet other cases no separate precursor is seen. An unusual feature of the precursor outburst of NN Cam is that it was apparently the same amplitude as the subsequent superoutburst, although this observation is based on a single data point.

This raises the question as to whether the normal outburst which NN Cam underwent in 2007 was a precursor that triggered the superoutburst. In other words are the two outbursts connected or are they separate events? One problem in addressing this question is that there is a gap of 6 days in the data between the last observation of the normal outburst and the detection of the superoutburst. In the case of VW Hyi, the maximum gap between the peak of the normal outburst and the detection of the subsequent superoutburst is ~ 5 days, whereas the maximum of the 2007 outburst of NN Cam occurred some 9 days before the detection of the superoutburst. However, we suspect that the superoutburst actually started a few days earlier since at the time the superoutburst was detected the star was already beginning to fade at a much faster rate than observed at the beginning of the 2009 superoutburst, which itself was detected near the onset of the superoutburst. SU UMa systems generally exhibit rather similar light curve profiles from one superoutburst to the next. This has, for example, been demonstrated in the case of VW Hyi over a multiplicity of superoutbursts [18] and recently the remarkable similarity of the light curves of the 2001 and 2009 superoutbursts of CP Dra has been shown [20]. Thus we attempted to combine the plots of the two superoutbursts of NN Cam by aligning the times of similar brightness during the plateau phase. The resulting light curve in Figure 9 shows reasonable alignment and leads us to suggest that the 2007 superoutburst may have been ongoing for 4 days before it was detected, in which case we missed the first one-third of the plateau phase. If this were the case then the time between the maximum of the normal outburst and the beginning of the superoutburst may have been as little as 2 days. Such an interval is similar to the intervals between precursor outburst and superoutburst in VW Hyi and leads us to propose that the normal outburst was in fact the event which triggered the superoutburst. We plot an example of a VW Hyi outburst light curve in Figure 10 in which the maximum of the precursor outburst occurs about 4 days before the detection of the superoutburst. We note the similarity of this light curve profile to NN Cam's 2007 outburst.

Rather few precursor outbursts in other SU UMa systems have been studied in detail. Thus the presence of modulations in the decline from the precursor outburst in 2007 is noteworthy, especially if they are related to $P_{orb}$ as we speculate they are. We note that in an independent analysis of the 2007 precursor, apparently based on the same observational data as in the present paper, Kato *et al.* [21] also reported modulations which were consistent with $P_{orb}$. This appears to be different from the situation in the likely precursor to the 2008 superoutburst, where no modulations were detected, at least during the time we observed the star. Orbital humps were present during the precursor to the 2008 superoutburst HS 0417+7445 [12]. However, no significant modulations were seen during the precursor outburst of the 2003 superoutburst in GO Com [13]. By contrast, Kato [16] analysed observations during the precursor outburst of the 1993 outburst of QZ Vir (=T Leo) and concluded that superhumps were already present during the decline from the precursor. Studies of precursor outbursts in other SU UMa systems, in particular regarding whether orbital or superhump modulations are present, may shed light on the mechanism by which a superoutburst is triggered. Hence, further studies of NN Cam during future outbursts are strongly encouraged. Based on the intervals between the three observed superoutbursts of 12.5 and 13.5 months, and assuming no superoutbursts were missed, superoutbursts appear to be an annual event.

Finally, we can also ask whether there was a precursor outburst before the 2009 superoutburst which we missed. As noted above, we caught the superoutburst very near the beginning. The top panel in Figure 4 shows that observations made 3 and 4 days before this superoutburst was detected showed that NN Cam was at or near quiescence. Moreover, we can also see from Figure 9, where the 2007 and 2009 outburst are plotted on the same timescale, that these quiescence observations were probably made at the same time as the precursor occurred in the 2007 outburst. We therefore consider a precursor to the 2009 superoutburst to have been unlikely. Unfortunately no observations were made in the 10 days prior to these quiescence observations to be certain of this. As noted above, both the presence and absence of precursor outbursts has been noted in VW Hyi at different times. Further observational coverage of NN Cam will throw light on how often precursor outbursts occur, although given its likely low outburst frequency, this is a long term project.

**Conclusions**

We report extensive photometry of the first confirmed superoutburst of NN Cam in 2007. It started with a normal outburst which lasted about 4 days and was the precursor to a superoutburst which lasted at least 13 days. Both the precursor and the superoutburst had an amplitude of 4.9 mag above mean quiescence. Superhumps with a maximum peak-to-peak amplitude of 0.22 mag were detected during the superoutburst. Analysis of our data reveals a superhump period $P_{sh}$ = 0.07385(56) d, although the superhump period decreased continuously with $dP_{sh}/dt$ = -1.72(23) x $10^{-3}$. We used our measurement to confirm that the shorter of two possible values of $P_{orb}$ reported by another researcher is the correct one, $P_{orb}$ = 0.0717 d. Modulations were detected during the decline from the precursor outburst which were consistent with $P_{orb}$.

The 2008 outburst was rather poorly observed, although we present evidence that this too was a superoutburst.

Observations of the 2009 outburst showed this was also a superoutburst, but in this case we could find no evidence for a precursor. Analysis of superhump times gave a constant superhump period with $P_{sh}$ = 0.07414(44) d.

We report a superhump period excess of $\varepsilon$ = 0.030(8) to 0.034(6), which is typical for SU UMa dwarf novae of similar orbital period, and estimate the binary mass ratio $q = M_{wd}/M_{sec}$ ≈ 0.11 to 0.17.

**Acknowledgements**

The authors gratefully acknowledge the use of observations from the AAVSO International Database contributed by observers worldwide. We are grateful to Professor Joe Patterson (Columbia University, USA) for allowing us to use data from the Center for Backyard Astrophysics (CBA) as well as for promoting the CBA observational campaign in 2007. We thank our referees, Dr. Boris Gänsicke (University or Warwick, UK) and Dr. Robert Smith (University of Sussex, UK) for helpful comments that have improved the paper. This research made use of SIMBAD and Vizier, operated through the Centre de Données Astronomiques de Strasbourg, and the NASA/Smithsonian Astrophysics Data System.

**Addresses**

JS: "Pemberton", School Lane, Bunbury, Tarporley, Cheshire, CW6 9NR, UK [bunburyobservatory@hotmail.com]
JF: Center for Backyard Astrophysics (Utah), 4175 E. Red Cliffs Drive, Kanab, UT 84741, USA [jfoote@scopecraft.com]
WMJ: 4587 Rockaway Loop, Rio Rancho, NM 871224, USA [mack-julian@cableone.net]
TK: CBA New Mexico, PO Box 1351 Cloudcroft, New Mexico 88317, USA [tom_krajci@tularosa.net]
IK: Vihorlat Observatory, Mierova 4, 06601 Humenne, Slovakia [vihorlatobs1@stonline.sk]
IM: Furzehill House, Ilston, Swansea, SA2 7LE, UK [furzehillobservatory@hotmail.com]
EM: Lauwin-Planque Observatory, F-59553 Lauwin-Planque, France [etmor@free.fr]
RS: 2336 Trailcrest Dr., Bozeman, MT 59718, USA [richard@theglobal.net]
IS: Department of Astronomy, Odessa National University, T.G.Shevchenko Park, Odessa, 65014, Ukraine [irina_solovyova@mail.ru]
BS: CBA Flanders, Patrick Mergan Observatory, Koningshofbaan 51, Hofstade, Aalst, Belgium [staels.bart.bvba@pandora.be]
TV, Center for Backyard Astrophysics (Belgium), Walhostraat 1A, B-3401 Landen, Belgium [tonny.vanmunster@cbabelgium.com]

| Observer | Telescope | CCD |
|---|---|---|
| Vanmunster | 0.35 m SCT | SBIG ST-7XME |
| Krajci 2007 | 0.28 m SCT | SBIG ST-7E |
| Krajci 2008 | 0.35 m SCT | SBIG ST9-XME |
| Solovieva & Kudzej | 0.28 m reflector | Meade DSI Pro |
| Foote | 0.6 m reflector | SBIG ST-8XE |
| Julian | 0.3 m SCT | SBIG ST10XME |
| Sabo | 0.43 m reflector | SBIG STL-1001 |
| Shears | 0.28 m SCT | Starlight Xpress SXVF-H9 |
| Staels | 0.28 m SCT | Starlight Xpress MX716 |
| Miller | 0.35 m SCT | Starlight Xpress SXVF-H16 |
| Morelle | 0.4 m SCT | Starlight Xpress SXV-M7 |

**Table 1: Equipment used**

| Start time | Duration (h) | Filter | Observer |
|---|---|---|---|
| 2454355.431 | 1.51 | C | Vanmunster |
| 2454356.494 | 1.32 | C | Vanmunster |
| 2454357.523 | 0.22 | C | Vanmunster |
| 2454363.517 | 2.47 | C | Vanmunster |
| 2454364.493 | 3.41 | V | Solovieva & Kudzej |
| 2454365.312 | 8.01 | C | Vanmunster |
| 2454365.492 | 0.13 | V | Solovieva & Kudzej |
| 2454365.753 | 6.07 | C | Krajci |
| 2454366.280 | 8.83 | C | Vanmunster |
| 2454366.493 | 3.45 | C | Solovieva & Kudzej |
| 2454367.545 | 2.25 | C | Solovieva & Kudzej |
| 2454368.358 | 5.04 | C | Vanmunster |
| 2454368.774 | 5.40 | C | Krajci |
| 2454369.494 | 3.43 | C | Solovieva & Kudzej |
| 2454369.783 | 5.18 | C | Foote |
| 2454370.783 | 5.18 | C | Foote |
| 2454376.775 | 5.37 | C | Foote |

**Table 2: Log of time-series observations, 2007 outburst**

| Start time   | Duration (h) | Filter | Observer |
|--------------|--------------|--------|----------|
| 2455137.658  | 7.20         | C      | Julian   |
| 2455138.637  | 6.79         | C      | Julian   |
| 2455139.735  | 0.31         | V      | Sabo     |
| 2455140.624  | 7.24         | V      | Sabo     |
| 2455142.271  | 2.78         | C      | Shears   |
| 2455143.236  | 9.79         | C      | Staels   |
| 2455143.430  | 0.35         | C      | Shears   |
| 2455144.224  | 3.16         | C      | Staels   |
| 2455146.386  | 4.24         | V      | Miller   |

**Table 3: Log of time-series observations, 2009 outburst**

| Start time   | Duration (h) | Filter | Observer |
|--------------|--------------|--------|----------|
| 2454759.300  | 10.24        | V      | Morelle  |
| 2454781.773  | 5.92         | V      | Krajci   |
| 2454782.773  | 3.50         | C      | Krajci   |
| 2454783.783  | 5.68         | C      | Krajci   |

**Table 4: Log of time-series observations, 2008 outburst**

| Superhump cycle | Superhump maximum (HJD) | O-C (days) | Uncertainty (d) | Superhump amplitude (mag) |
|---|---|---|---|---|
| 0 | 2454363.5481 | -0.0094 | 0.0018 | 0.14 |
| 13 | 2454364.5148 | -0.0027 | 0.0008 | 0.15 |
| 14 | 2454364.5850 | -0.0064 | 0.0007 | 0.16 |
| 24 | 2454365.3307 | 0.0009 | 0.0008 | 0.16 |
| 25 | 2454365.4032 | -0.0006 | 0.0010 | 0.19 |
| 26 | 2454365.4773 | -0.0003 | 0.0008 | 0.16 |
| 27 | 2454365.5537 | 0.0023 | 0.0005 | 0.22 |
| 28 | 2454365.6300 | 0.0048 | 0.0005 | 0.21 |
| 30 | 2454365.7766 | 0.0036 | 0.0009 | 0.18 |
| 31 | 2454365.8480 | 0.0011 | 0.0005 | 0.14 |
| 32 | 2454365.9231 | 0.0024 | 0.0013 | 0.14 |
| 33 | 2454365.9965 | 0.0019 | 0.0006 | 0.16 |
| 39 | 2454366.4386 | 0.0009 | 0.0009 | 0.20 |
| 40 | 2454366.5164 | 0.0049 | 0.0008 | 0.18 |
| 41 | 2454366.5871 | 0.0017 | 0.0003 | 0.20 |
| 55 | 2454367.6198 | 0.0006 | 0.0004 | 0.18 |
| 66 | 2454368.4334 | 0.0018 | 0.0009 | 0.15 |
| 67 | 2454368.5065 | 0.0011 | 0.0007 | 0.17 |
| 71 | 2454368.8012 | 0.0003 | 0.0006 | 0.14 |
| 72 | 2454368.8754 | 0.0008 | 0.0013 | 0.14 |
| 73 | 2454368.9484 | -0.0001 | 0.0008 | 0.13 |
| 81 | 2454369.5355 | -0.0038 | 0.0007 | 0.12 |
| 82 | 2454369.6125 | -0.0006 | 0.0015 | 0.13 |
| 85 | 2454369.8318 | -0.0029 | 0.0018 | 0.11 |
| 86 | 2454369.9052 | -0.0034 | 0.0018 | 0.08 |
| 87 | 2454369.9785 | -0.0039 | 0.0007 | 0.09 |
| 99 | 2454370.8640 | -0.0046 | 0.0007 | 0.15 |
| 100 | 2454370.9333 | -0.0092 | 0.0008 | 0.12 |

**Table 5: Superhump maximum times during 2007 outburst**

| Superhump cycle | Superhump maximum (HJD) | O-C (days) | Uncertainty (d) | Superhump amplitude (mag) |
|---:|---:|---:|---:|---:|
| 0 | 2455138.6756 | -0.0010 | 0.0007 | 0.08 |
| 1 | 2455138.7490 | -0.0017 | 0.0013 | 0.11 |
| 2 | 2455138.8234 | -0.0015 | 0.0009 | 0.12 |
| 3 | 2455138.8963 | -0.0027 | 0.0008 | 0.14 |
| 27 | 2455140.6820 | 0.0036 | 0.0007 | 0.26 |
| 28 | 2455140.7565 | 0.0040 | 0.0008 | 0.22 |
| 29 | 2455140.8304 | 0.0038 | 0.0012 | 0.24 |
| 30 | 2455140.9043 | 0.0035 | 0.0008 | 0.22 |
| 49 | 2455142.3086 | -0.0009 | 0.0013 | 0.14 |
| 62 | 2455143.2715 | -0.0017 | 0.0001 | 0.16 |
| 63 | 2455143.3460 | -0.0015 | 0.0003 | 0.15 |
| 64 | 2455143.4208 | -0.0008 | 0.0003 | 0.15 |
| 65 | 2455143.4911 | -0.0046 | 0.0005 | 0.16 |
| 66 | 2455143.5645 | -0.0054 | 0.0004 | 0.16 |
| 75 | 2455144.2382 | 0.0011 | 0.0022 | 0.17 |
| 76 | 2455144.3122 | 0.0010 | 0.0028 | 0.18 |
| 106 | 2455146.5371 | 0.0016 | 0.0013 | 0.15 |

**Table 6: Superhump maximum times during 2009 outburst**

| Group 1 | | Group 2 | |
|---|---|---|---|
| Frequency (cycles/day) | Period (d) | Frequency (cycles/day) | Period (d) |
| 13.970(88) | 0.0716(5) | 13.480(117) | 0.0742(6) |
| 12.970(76) | 0.0771(4) | 12.478(95) | 0.0801(5) |
| 14.970(79) | 0.0668(4) | 14.483(79) | 0.069(4) |

**Table 7: Main signals in the power spectrum shown in Figure 8**

Signal strength decreases down each column. In Figure 8, Group 1 signals are identified by a red bar and Group 2 signals by a blue bar

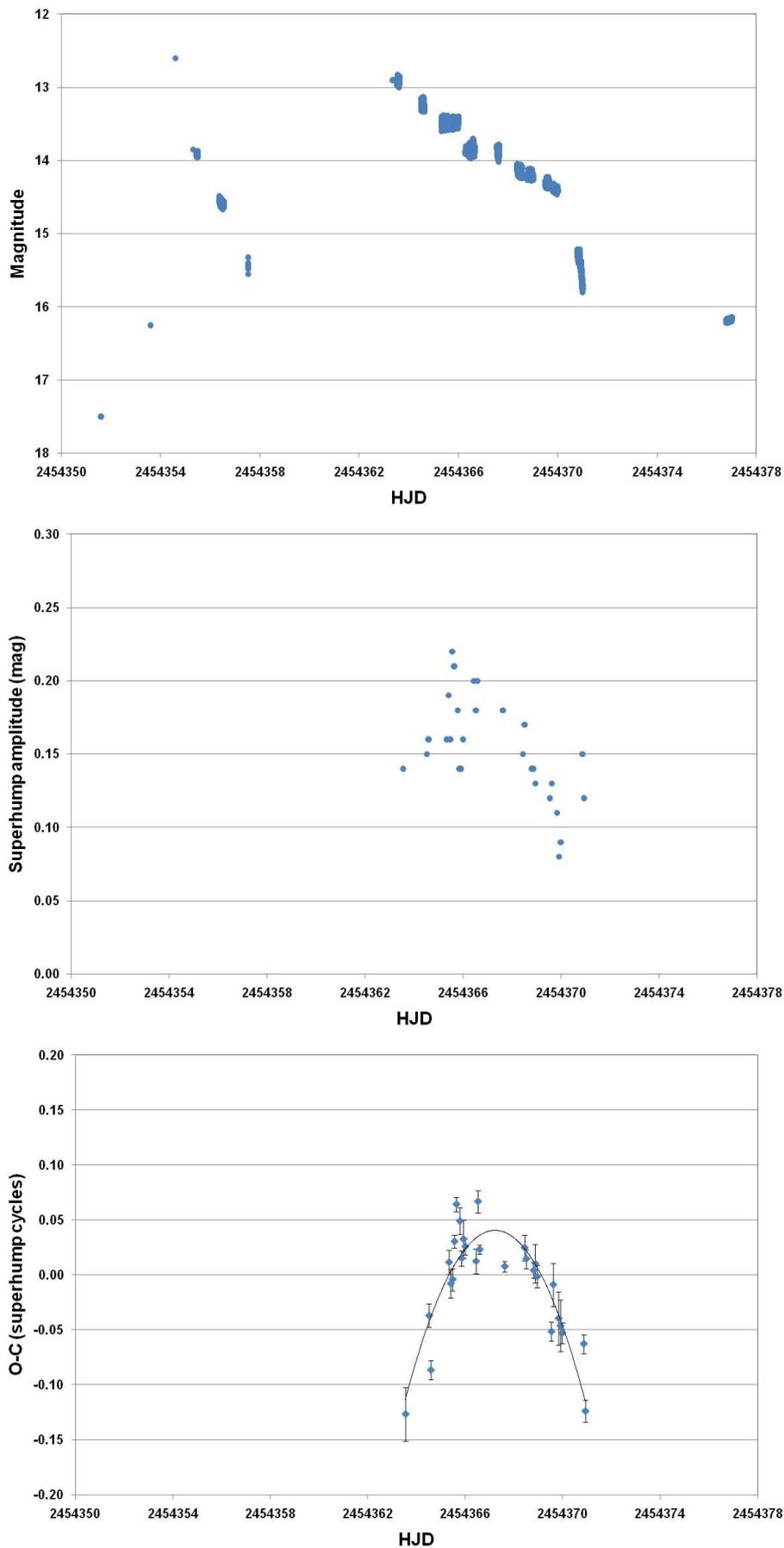

**Figure 1: 2007 outburst**
Top: outburst light curve. Middle: superhump amplitude. Bottom: O-C of superhump residuals. The solid line represents a quadratic fit to the data

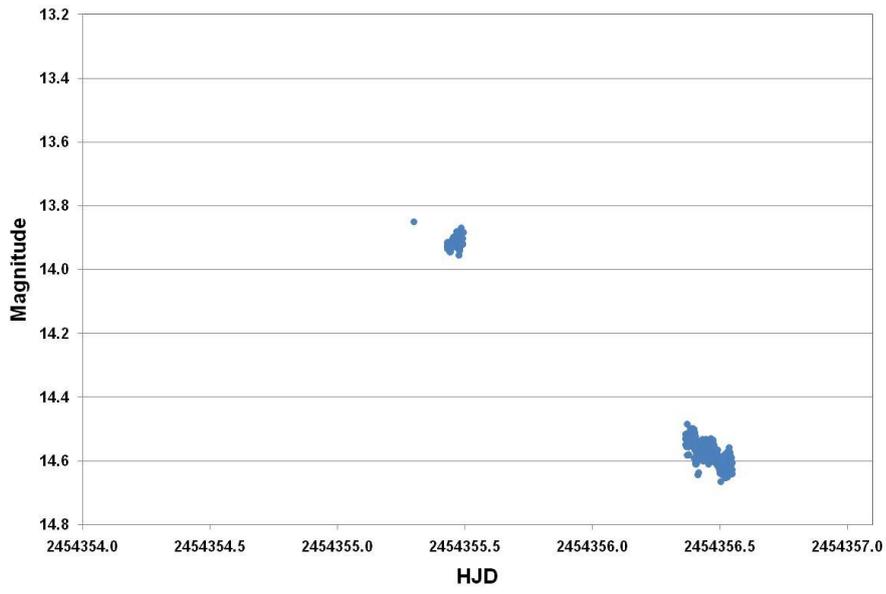

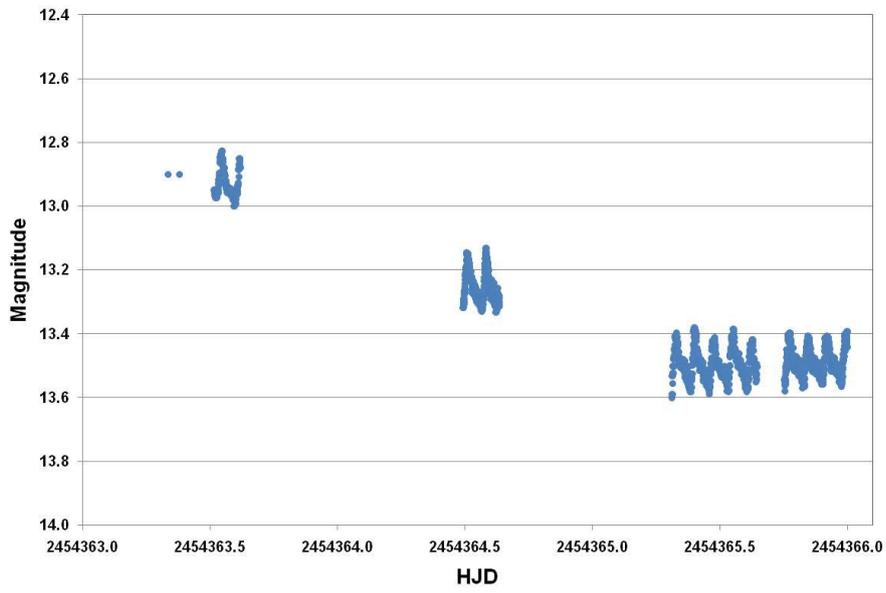

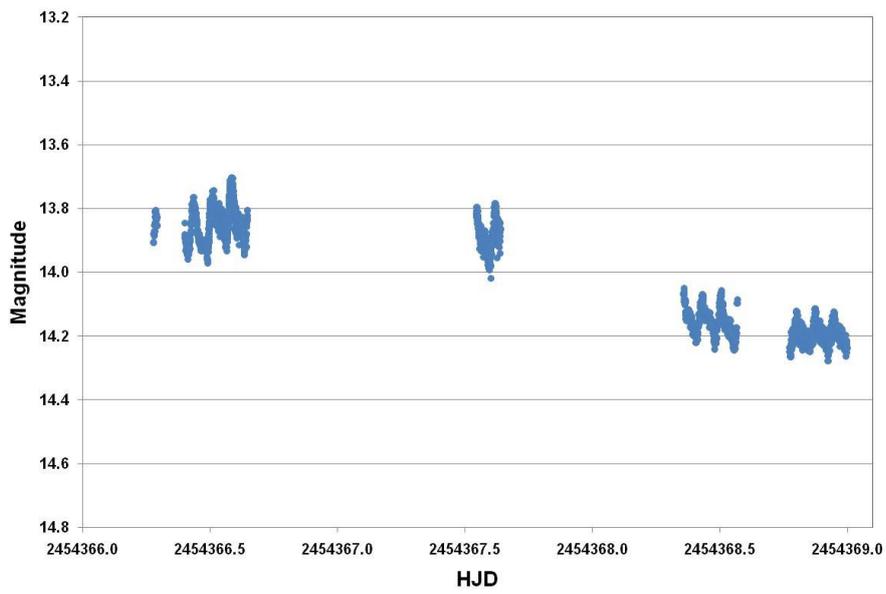

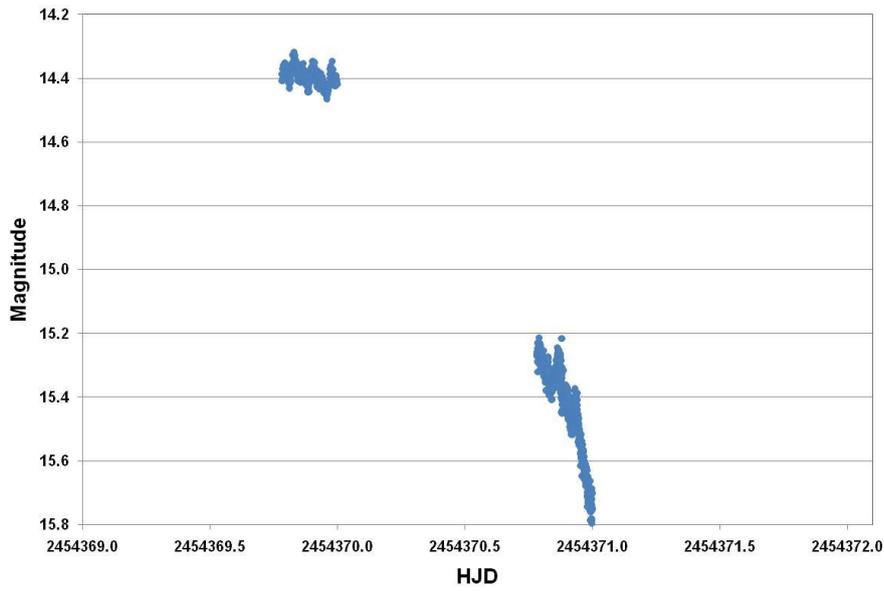

**Figure 2: Photometry from the 2007 outburst**

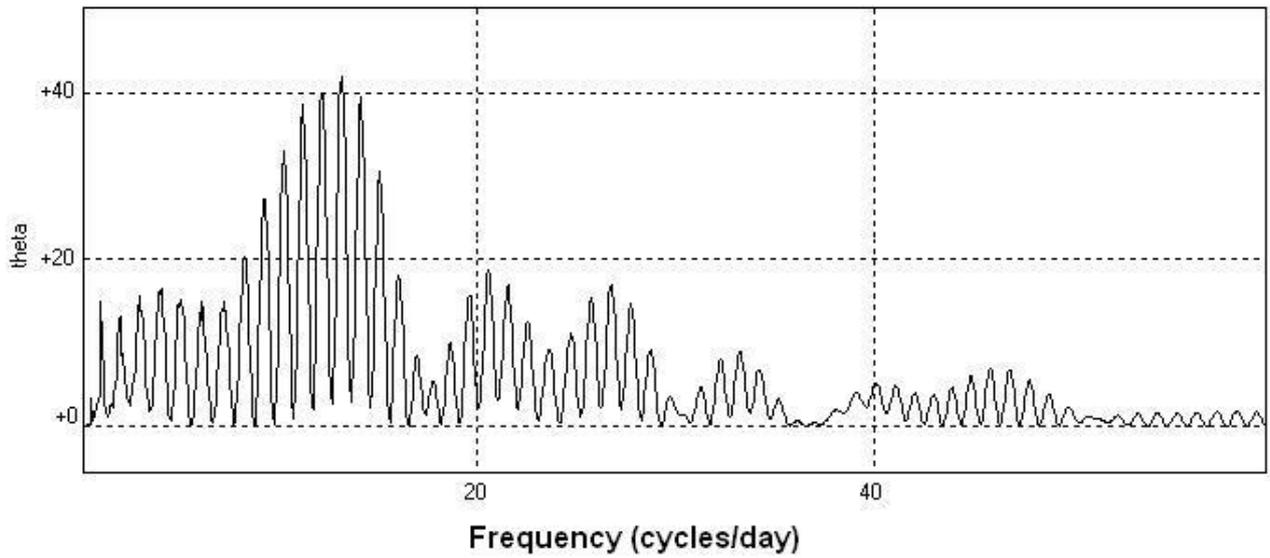

**Figure 3: Power spectrum of the data from the 2007 normal outburst (JD 2454356 to 2454358)**

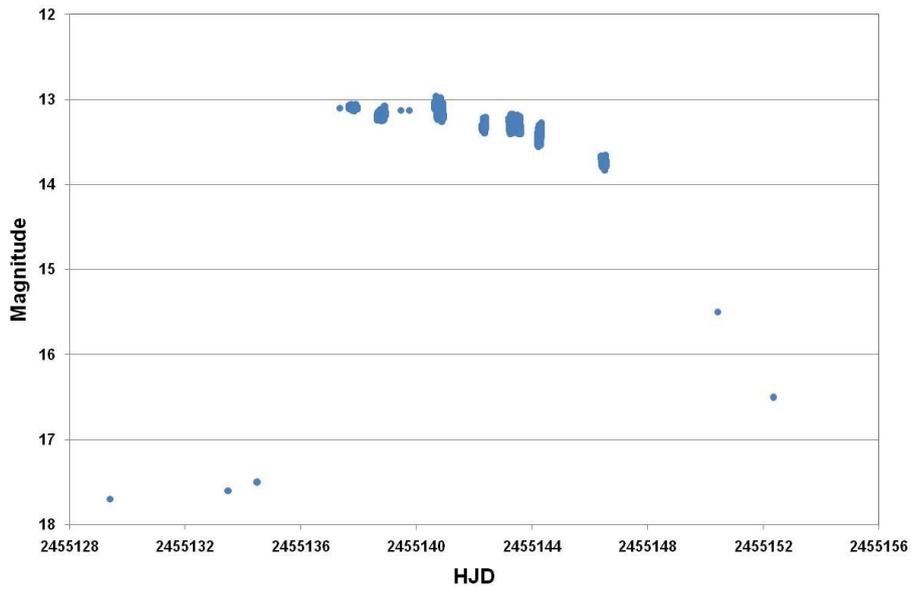

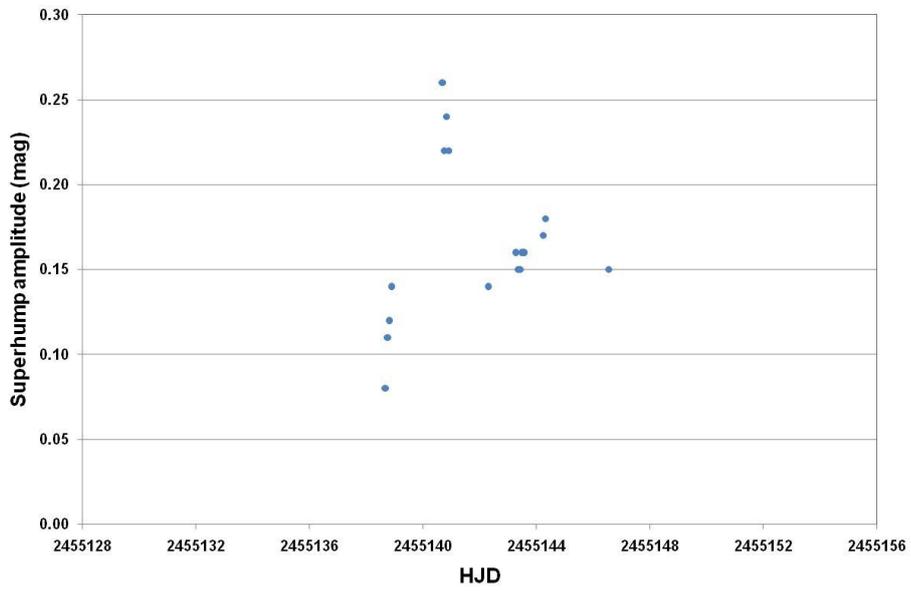

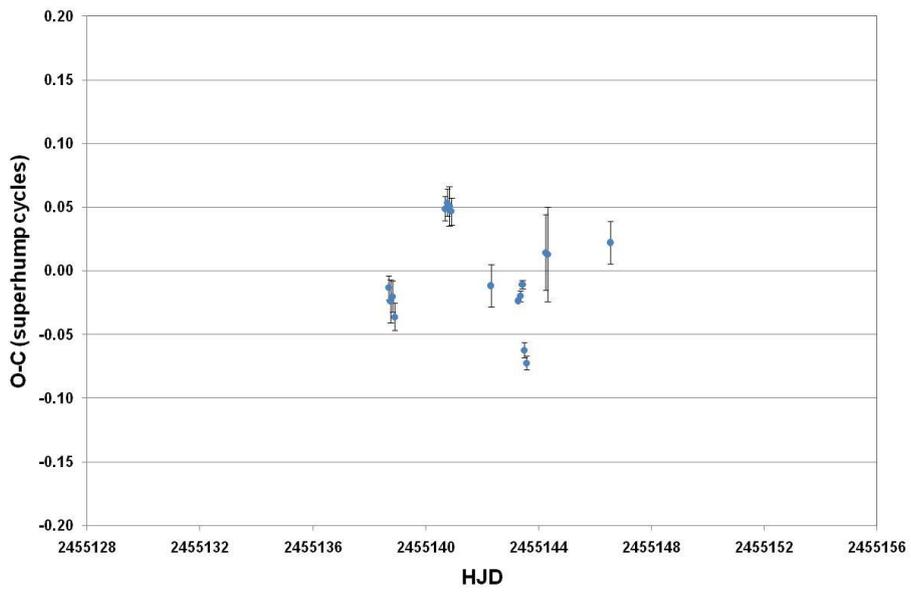

**Figure 4: 2009 outburst**
**Top: outburst light curve. Middle: superhump amplitude. Bottom: O-C of superhump residuals**

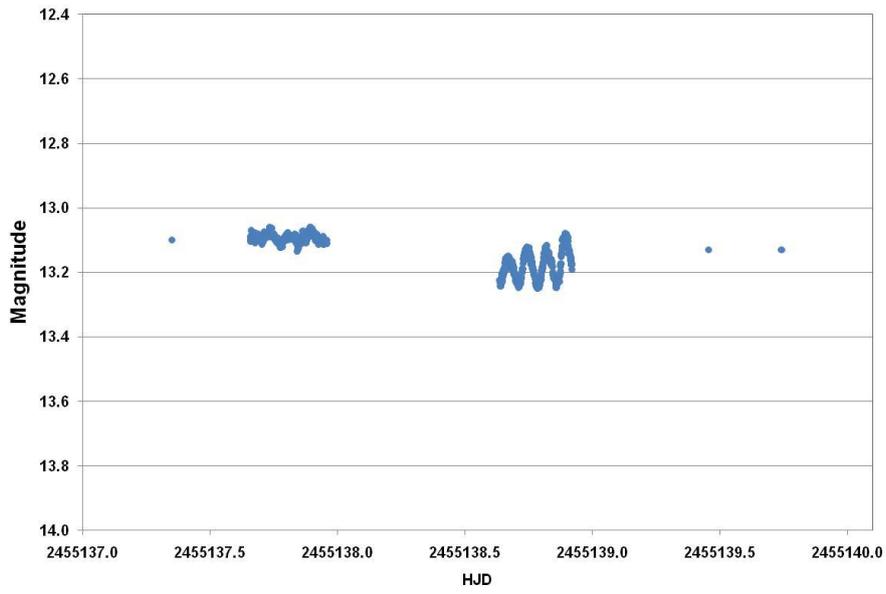
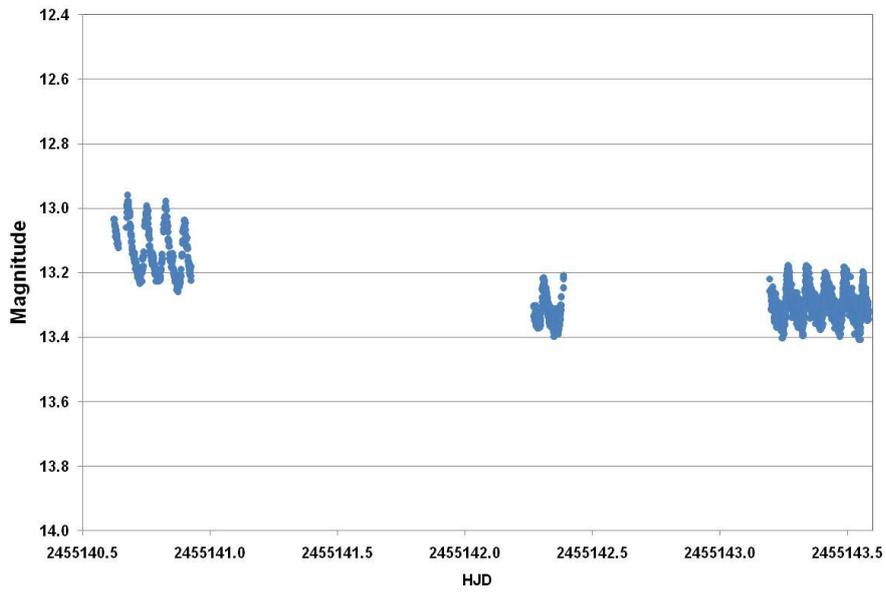
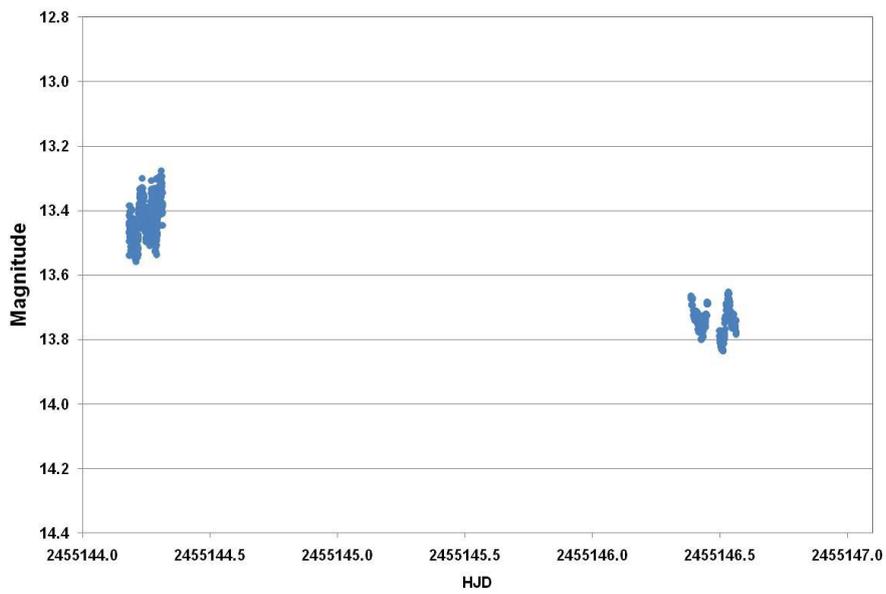

**Figure 5: Photometry from the 2009 outburst**

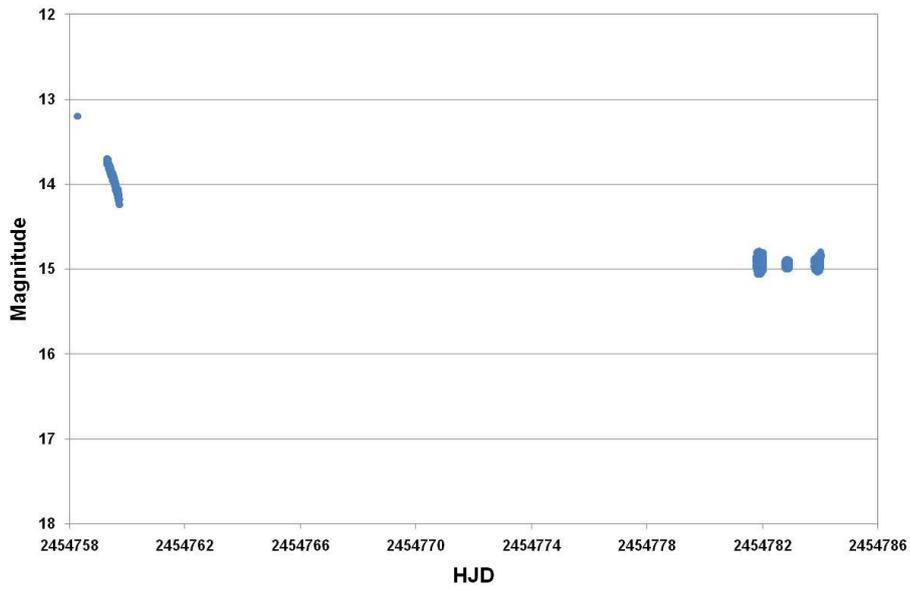

**Figure 6: Light curve of the 2008 outburst**

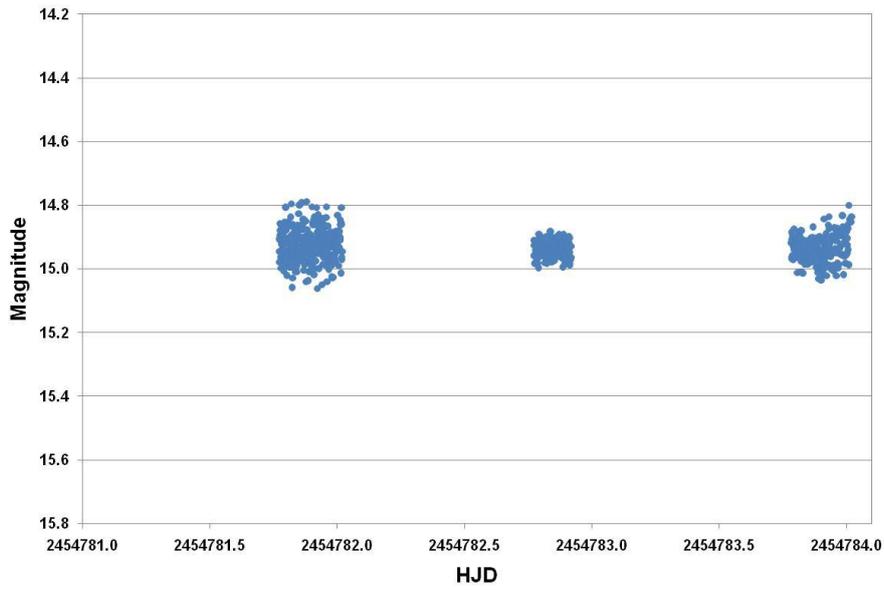

**Figure 7: Photometry from the 2008 outburst**

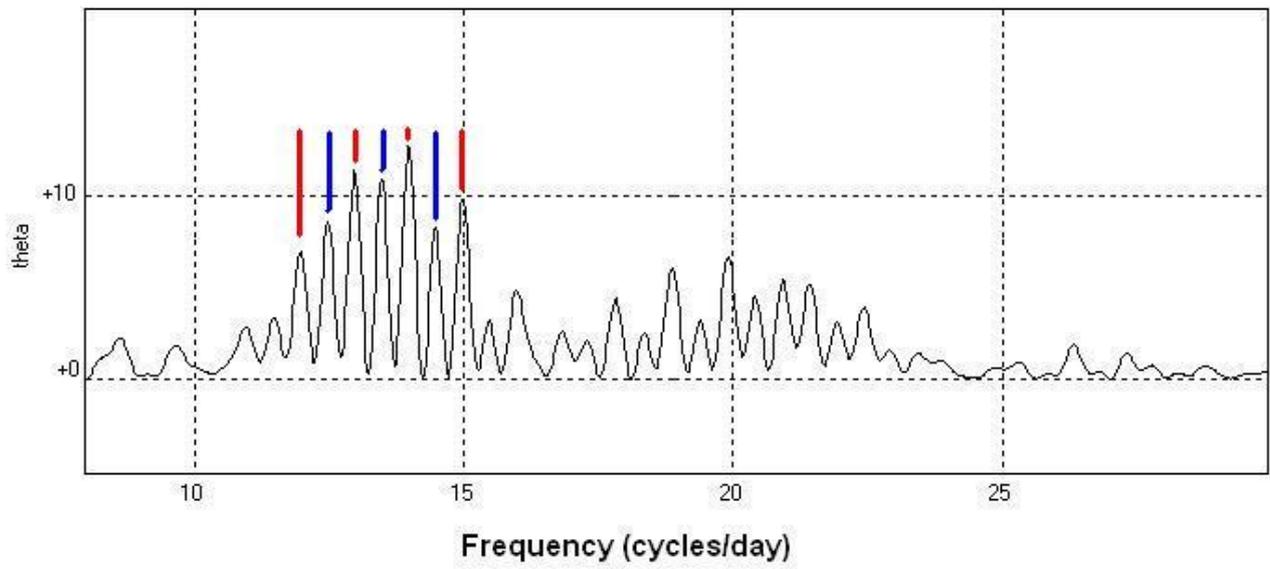

**Figure 8: Power spectrum of the 2008 data (JD 2454781 to 2454784)**

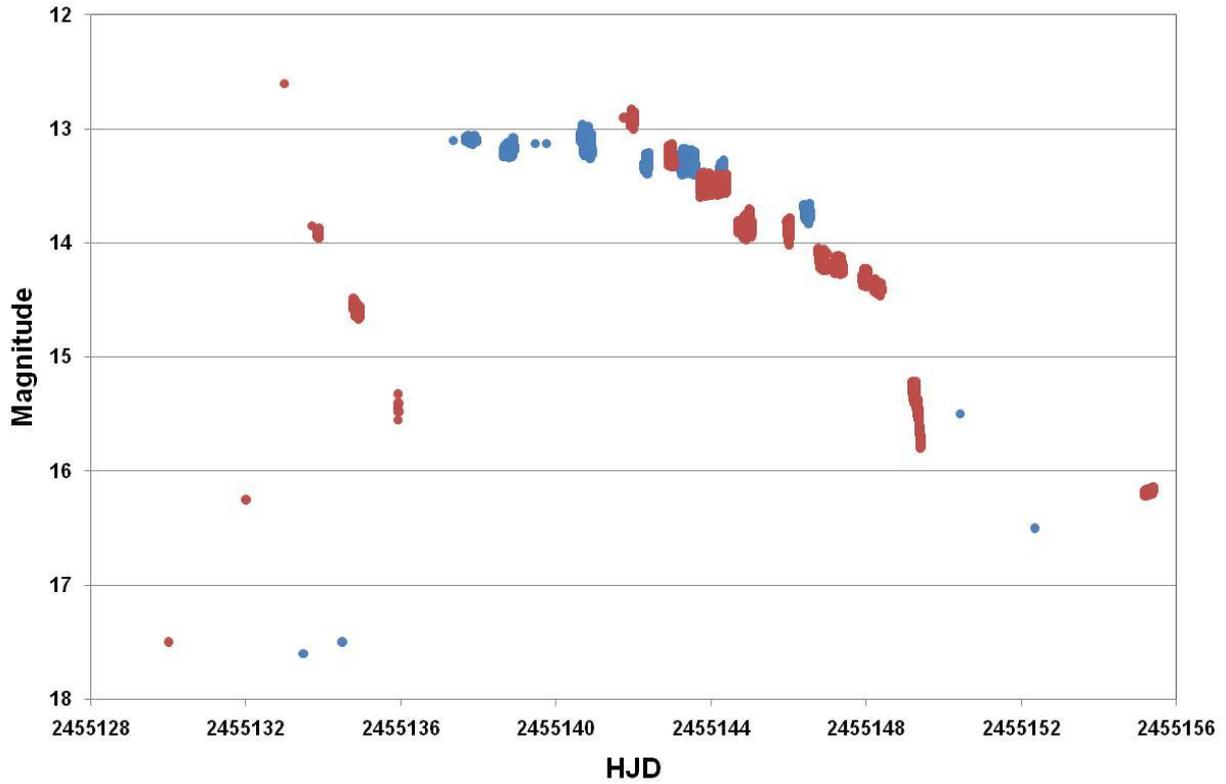

**Figure 9: Combined light curves of the 2007 and 2009 outbursts**
2007 data: red, 2009 data: blue. 2007 data were transformed by adding JD 778.4

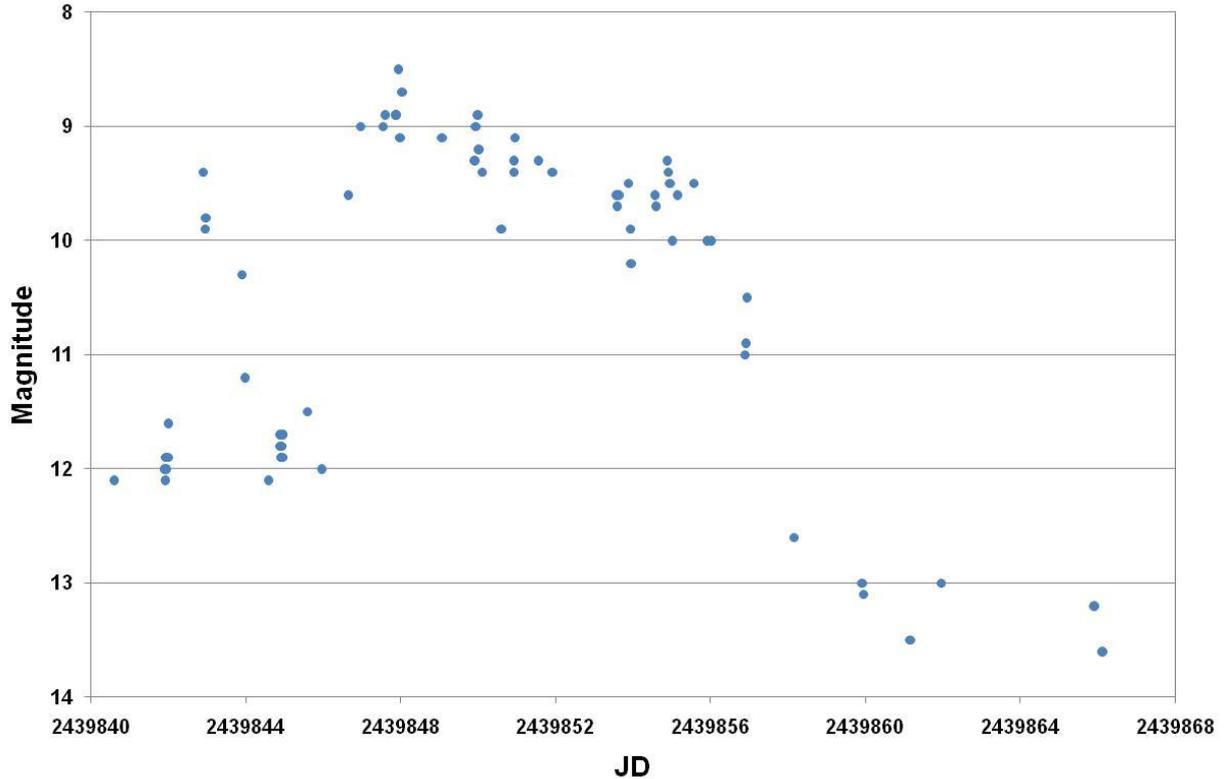

**Figure 10: Example of VW Hyi outburst showing a distinct precursor and a superoutburst**
Data from the AAVSO International Database